# Evaluation of Elekta Symmetry™ 4D IGRT system by using moving lung phantom

**Hun-Joo Shin, Shin-Wook Kim, and Chul Seung Kay**

*Department of Radiation Oncology, Incheon St. Mary's Hospital, College of Medicine, The Catholic University of Korea, Incheon 403-720, Korea*

**Jae-Hyuk Seo**

*Department of Radiation Oncology, Bucheon St. Mary's hospital, College of Medicine, The Catholic University of Korea, Bucheon 420-717, Korea*

**Gi-Woong Lee**

*Department of Medical Physics, Kyonggi University of Korea, School of Medicine, Suwon, 443-760, Korea*

**Ki-Mun Kang**

*Department of Radiation Oncology, Gyeong-Sang National University Hospital, Chinju 660-702, Korea*

**Hong Seok Jang and Young-nam Kang***

*Department of Radiation Oncology, Seoul St. Mary's hospital, College of Medicine, The Catholic University of Korea, Seoul 135-701, Korea*

**Purpose:** 4D CBCT is a beneficial tool for the treatment of movable tumors, because it can help us to understand where the tumors are actually located and have a precise treatment plan. However, there is a limitation that general CBCT images cannot perfectly help the sophisticated registration. On the other hand, Symmetry[TM] 4D IGRT system of Elekta can offer the 4D CBCT registration option. In this study, we intend to evaluate the usefulness of Symmetry[TM].

**Method and Materials:** Planning CT images of the CIRS moving lung phantom were acquired from 4D MDCT. And they are sorted as 10 phases from 0% phase to 90% phase. The thickness of CT images was 1 mm. Acquired MDCT images were transferred to the contouring software and a virtual target was generated. An one arc VMAT plan was performed by using the treatment planning system on the virtual target. Finally, the movement of the phantom was verified through XVI Symmetry[TM] system.

**Results:** The physical movement of CIRS moving lung phantom was ± 10.0 mm in superior-inferior direction, ± 1.0 mm in lateral direction, and ± 2.5 mm in anterior-posterior direction. The movement of the phantom was measured from 4D MDCT registration as ± 10.2 mm in superior-inferior direction, ± 0.9 mm in lateral direction, and ± 2.45 mm in anterior-posterior direction. The movement of the phantom was measured from Symmetry[TM] registration as ± 10.1 mm in superior-inferior direction, ± 0.9 mm in lateral direction, and ± 2.4 mm in anterior/posterior direction.

**Conclusion:** It is confirmed that 4D CBCT is a beneficial tool for the treatment of movable tumors. Therefore, 4D registration of Symmetry[TM] can increase the precision of the registration when a movable tumor is a target of radiation treatment.

Email: ynkang33@gmail.com

Fax: +82-2-2258-2532

# I. INTRODUCTION

Reproducibility of patients' setup and accuracy of target localization are improved in modern radiation therapy through the image guided radiation therapy (IGRT) technique. For these purposes, cone beam computed tomography (CBCT) is widely applied to the medical linear accelerators for the radiation therapy.[1] In the case of tumors moving with the respiration such as lung cancer, internal target volume (ITV) is determined by analysis of the four-dimensional multi detector computed tomography (4D MDCT) images. And the ranges of ITV are decided for the radiation treatment plans. Furthermore, thorough target localization procedures should be performed by using CBCT.[2-4] During CBCT image acquisition, however, the target motion caused by the patients' respiration may lead the motion artifacts. And it deteriorates the accuracy of the target localization. This might reduce the accuracy of target localization.[5]

Therefore, these problems can be minimized with the use of 4D CBCT reducing the ITV discrepancies between treatment and plan steps. Recently, many authors present the CBCT systems enable the respiratory correlation. Moreover, 4D CBCT used for the determination of planning target volume (PTV) margins has been studied in the state of art radiation therapy techniques including the adaptive IGRT and volumetric modulated arc therapy (VMAT).[6-10]

The aim of this study is to evaluate the usefulness of 4D CBCT system by using the 4D moving lung phantom.

## II. Material and Methods

*Modality and Radiation Treatment System*

In this study, the medical linear accelerator, Infinity$^{TM}$ (Elekta Oncology Systems, Crawley, West Sussex, UK) was used. XVI Symmetry$^{TM}$ 4.5.4.x version (Elekta Oncology Systems, Crawley, West Sussex, UK) was used. And the condition of image acquisition was 120 kVp, 422~4 mAs, and 1320 slice frames. The kilo-Voltage (kV) CBCT system with a kV X-ray tube and a flat-panel on the both side of gantry is attached. The active area of the flat-panel detector is $410 \times 410$ mm$^2$ and the pixel size of the image acquisition matrix is $512 \times 512$ pixels. Small mode of 4D CBCT scanning, the detector is symmetrically placed in the center of the field of view (FOV), was applied. The FOV was $270 \times 260$ mm$^2$. And the total 200 degree gantry angle (from 180 degree to 20 degree, on clockwise rotation) was set. For the treatment planning, Brilliance$^{TM}$ CT scanner (Philips Healthcare, Cleveland, OH) was used. The acquired CT images are sorted into 10 distinct phases ranging from full inhale (0%) to full exhale (50%) and back to inhale. The 1 mm slice thickness MDCT images are acquired. Acquired 4D MDCT images were transferred to the MIM 6.1.7 version (MIM software Inc. , Cleveland, OH), contouring software. After contouring the virtual target, 1 arc VMAT plan was performed from MONACO version 3.30.x (Elekta Oncology Systems, Crawley, West Sussex, UK), the treatment planning system. Then, the movement of the phantom was verified by using XVI Symmetry$^{TM}$ system.

*4D moving lung phantom*

The dynamic thorax phantom (CIRS Inc., Norfolk, VA) was used. This phantom mimics the humans' chest shapes and has a sphere tumor target in artificial lung material. This sphere tumor target can be moved in 3-dimensional directions. For this study, this sphere target was set to move up to ± 2.5 mm in anterior-posterior direction, ± 1 mm in lateral direction, and ± 10 mm in superior-inferior direction. The moving range of this dynamic thorax phantom was performed and that range of the motion in this study was set to closely mimic the patients' respiratory.[11] In figure 1 (a), experimental setup for the 4D

MDCT image acquisition of the 4D moving lung phantom is shown. The axial, coronal, and sagittal images are presented in figure 1 (b), (c), and (d). The experimental setup for the 4D CBCT image acquisition of the 4D moving lung phantom is shown in figure 2 (a). The image registration procedures after 4D CBCT image acquisition is shown in figure 2 (b). In figure 2 (c), it is shown that the sphere target movements after image registration were measured at each respiratory phase

## III. Results

While the 4D moving lung phantom is operating, sphere target movements were analyzed in both 4D MDCT and 4D CBCT cases. The amount of movements was compared with the pre-set amount of movements. The differences of movements are shown in table 1 and figure 3. On visual inspection, the physical movement of CIRS 4D moving lung phantom was verified. The movement of the phantom was measured from 4D MDCT registration as ± 10.2 mm in superior-inferior direction, ± 0.9 mm in lateral direction, and ± 2.45 mm in anterior-posterior direction. The movement of the phantom was measured from Symmetry$^{TM}$ registration as ± 10.1 mm in superior-inferior direction, ± 0.9 mm in lateral direction, and ± 2.4 mm in anterior/posterior direction.

By using the 4D moving lung phantom, the usefulness of 4D CBCT system for tumors moving with the respiration were evaluated. The differences between the movements analyzed from 4D MDCT and the pre-set movements were 0.2 mm (2%) in superior-inferior direction, 0.1 mm (10%) in lateral direction, and 0.05 mm (2%) in anterior-posterior direction. Furthermore, the differences between the movements analyzed from 4D CBCT and the pre-set movements were 0.1 mm (1%) in superior-inferior direction, 0.1 mm (10%) in lateral direction, and 0.1 mm (4%) in anterior-posterior direction.

## IV. Discussion and conclusion

The physical movement of CIRS 4D moving lung phantom was performed as we set in this study. The absolute values of measurement error were within 0.5 mm at all directions in both 4D MDCT and 4D CBCT cases. For instance, the maximum measurement error was ± 0.2 mm in superior-inferior direction on 4D MDCT and the minimum measurement error was ± 0.05 mm in anterior-posterior direction on 4D MDCT. However, the relative measurement error in lateral direction were 10% in both

4D MDCT and 4D CBCT cases, and the minimum measurement error in superior-inferior direction was 1 % on 4D CBCT. This is because that the pre-set motion range of 4D moving lung phantom in lateral direction was relatively shorter than in superior-inferior direction. If the motion range of 4D moving lung phantom is increased, the relative measurement error would be decreased. Due to the constant moving pattern of 4D moving lung phantom, the measurement error was equal in both positive and negative directions.

The 4D MDCT images of organs moving with the respiration have been widely used for the image registration in clinical practice. This is because the amount of measurement error is sufficiently less. According to the results of this study, 4D CBCT images also can be used as a method of the image registration with the less measurement error. If further studies with the variable target size, movement, shape, and respiratory pattern are performed, more accurate evaluation of 4D CBCT can be provided. Moreover, more accurate target localization through 4D CBCT images might be necessary to improve the effectiveness of the radiation treatment, especially in the case of tumors moving with the respiration such as lung cancer.

## ACKNOWLEDGEMENT

We would like to acknowledge the financial support from the R&D Convergence Program of MSIP (Ministry of Science, ICT and Future Planning) and ISTK (Korea Research Council for Industrial Science and Technology) of Republic of Korea (B551179-12-08-00, Development of Convergent Radio Therapy Equipment with O-arm CT)

**REREFENCES**

[1]  D. A. Jaffray, J. H. Siewerdsen, J. W. Wong and A. A. Martinez, Int. J. Radiat. Oncol. Biol. Phys. **53**, 1337 (2002).

[2]  J. P. Bissonnette, T. G. Purdie, J. A. Higgins, W. Li and A. Bezjak, Int. J. Radiat. Oncol. Biol. Phys. **73**, 927 (2009).

[3]  A. R. Yeung, J. G. Li, W. Shi, H. E. Newlin, A. Chvetsov, C. Liu, J. R. Palta and K. Olivier, Int. J. Radiat. Oncol. Biol. Phys. **74**, 1100 (2009).

[4]  Z. Wang, Q. J. Wu, L. B. Marks, N. Larrier and F. F. Yin, Int. J. Radiat. Oncol. Biol. Phys. **69**, 1618 (2007).

[5]  I. Vergalasova, J. Maurer and F. F. Yin, Med. Phys. **38**, 4689 (2011).

[6]  J. J. Sonke, L. Zijp, P. Remeijer and M. van Herk, Med. Phys. **32**, 1176 (2005).

[7]  A. Harsolia, G. D. Hugo, L. L. Kestin, I. S. Grills and D. Yan, Int. J. Radiat. Oncol. Biol. Phys. **70**, 582 (2008).

[8]  S. Kida, Y. Masutani, H. Yamashita, T. Imae, T. Matsuura, N. Saotome, K. Ohtomo, K. Nakagawa and A. Haga, Radiol Phys Technol **5**, 138 (2012).

[9]  K. Nakagawa, A. Haga, S. Kida, Y. Masutani, H. Yamashita, W. Takahashi, A. Sakumi, N. Saotome, T. Shiraki, K. Ohtomo, Y. Iwai and K. Yoda, J Radiat Res **54**, 152 (2013).

[10]  W. Takahashi, H. Yamashita, S. Kida, Y. Masutani, A. Sakumi, K. Ohtomo, K. Nakagawa and A. Haga, Int. J. Radiat. Oncol. Biol. Phys. **86**, 426 (2013).

[11]  J. A. Tanyi, M. Fuss, V. Varchena, J. L. Lancaster and B. J. Salter, Radiat. Oncol. **2**, 10 (2007).

Table 1. Movement and difference of 4D moving lung phantom

| | superior - inferior | | lateral | | anterior - posterior | |
|---|---|---|---|---|---|---|
| | Movement (mm) | Difference | Movement (mm) | Difference | Movement (mm) | Difference |
| **physical** | ± 10.0 | | ± 1.0 | | ± 2.5 | |
| **4D MDCT** | ± 10.2 | 2% | ± 0.9 | 10% | ± 2.45 | 2% |
| **4D CBCT** | ± 10.1 | 1% | ± 0.9 | 10% | ± 2.4 | 4% |

Figure Captions.

Figure 1. The experimental setup for the 4D moving lung phantom on the 4D MDCT images. 1(a) experimental setup for the 4D MDCT image acquisition of the 4D moving lung phantom and (b), (c), and (d) are the axial, coronal, and sagittal images of 4D MDCT .

Figure 2. The experimental setup for the 4D CBCT moving lung phantom on the 4D MDCT images and image registration procedures. (a) The experimental setup for the 4D CBCT image acquisition of the 4D moving lung phantom. (b) the image registration procedures after 4D CBCT image acquisition. (c) the sphere target movements after image registration were measured at each respiratory phase.

Figure 3. The illustration of sphere tumor target movements of the 4D moving lung phantom. (a) pre-set movements, analyzed movements from 4D MDCT, and analyzed movements from 4D CBCT. (b) the differences between analyzed movements from 4D MDCT and 4DCBCT with the comparisons of pre-set movements.

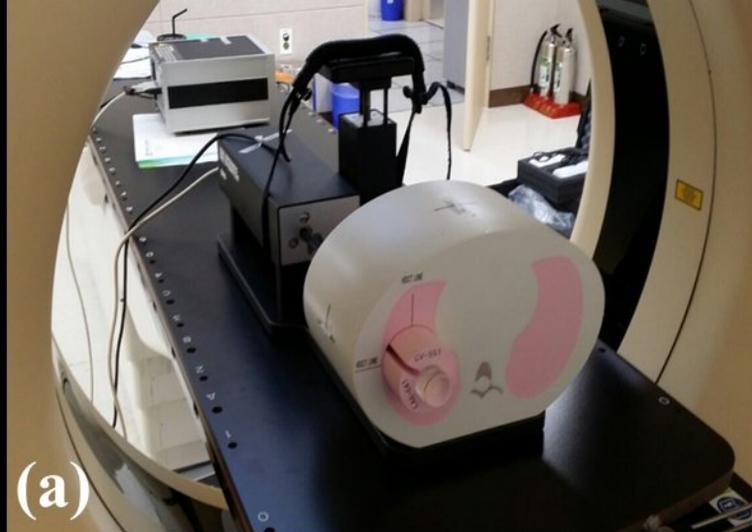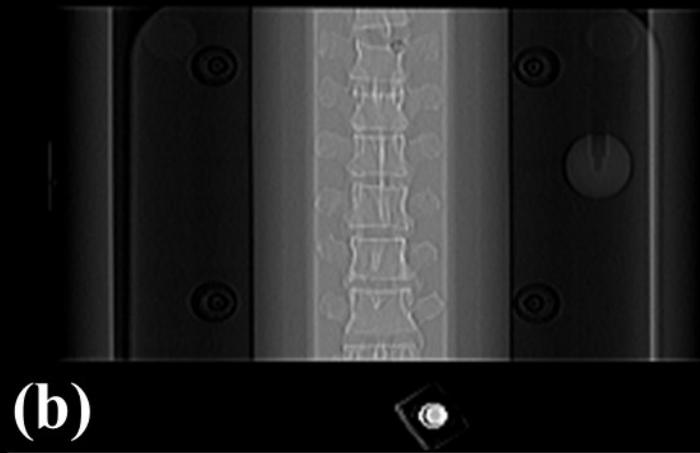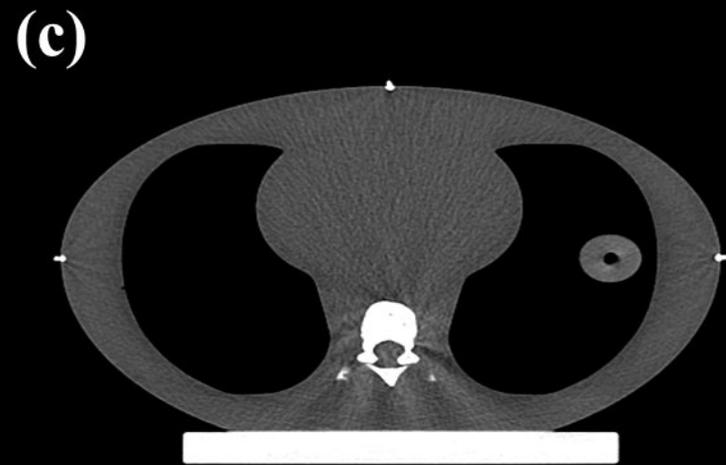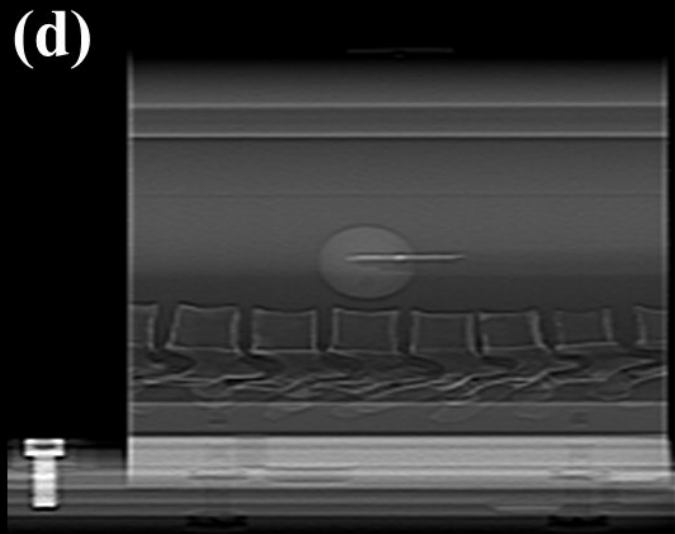

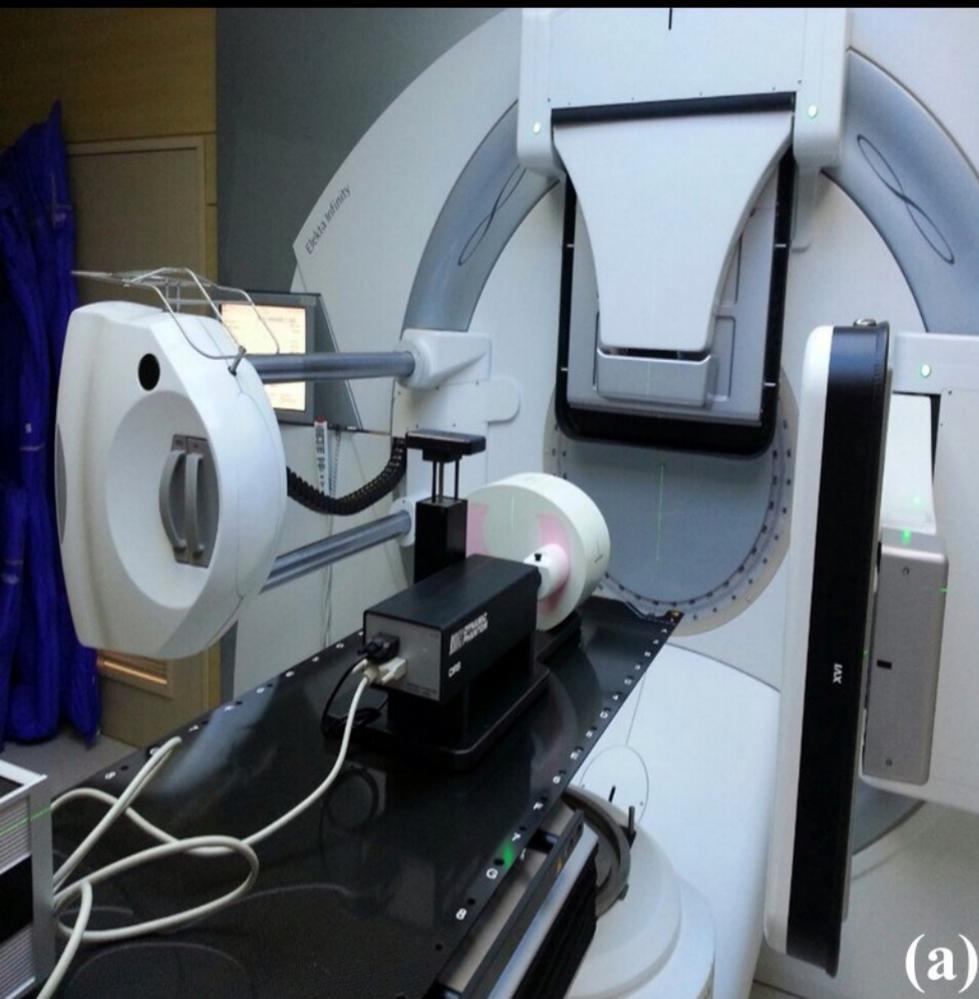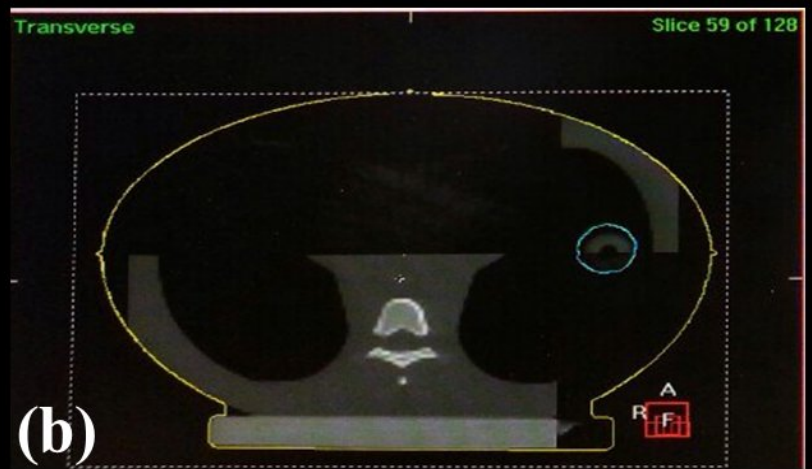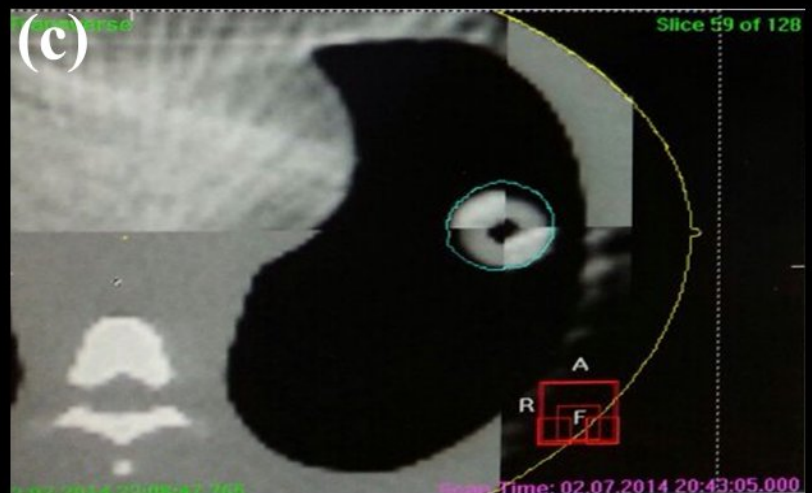

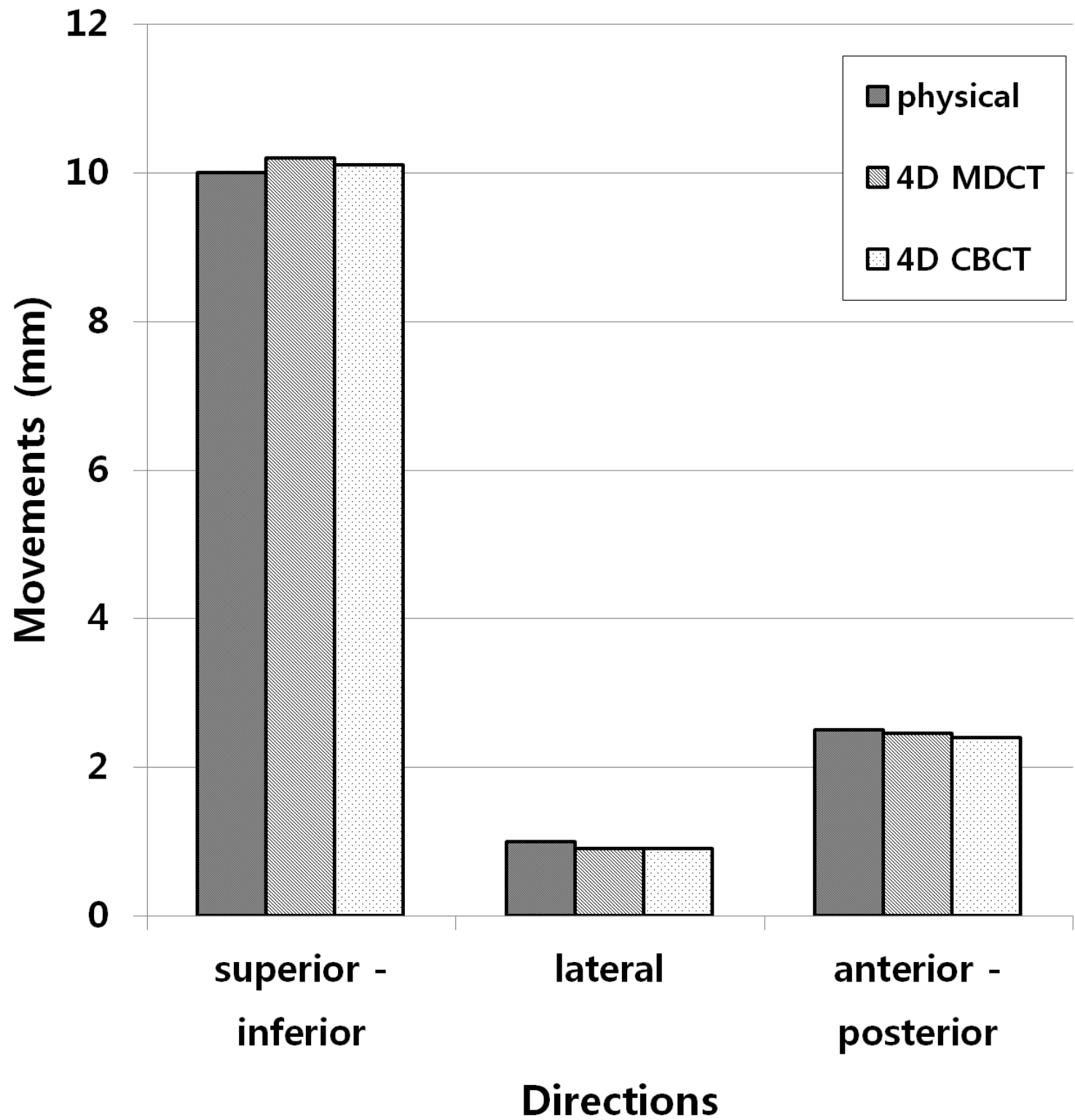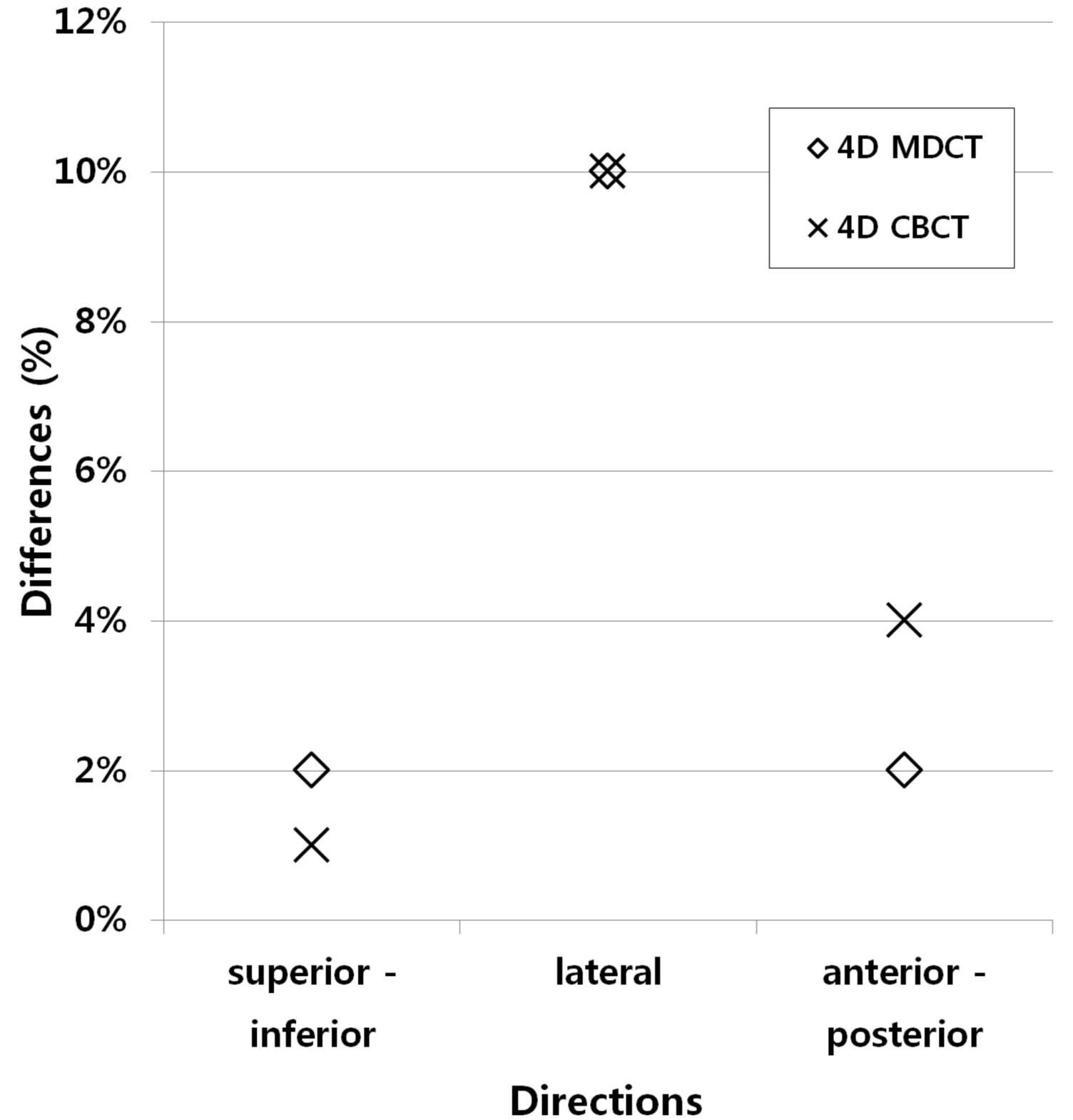